\documentstyle[multicol,epsf,aps,prl]{revtex}
\def\e{\varepsilon}
\begin{document}
\bibliographystyle{prsty}
\title{\Large\bf Kondo model for the ``0.7 anomaly" in transport through
a quantum point contact}
\author{Yigal Meir$^{1,2,\dagger}$, Kenji Hirose$^{3}$, and Ned S. Wingreen$^{1}$}
\address{
$^{1}$ NEC Research Institute, 4 Independence Way, Princeton, New Jersey
08540\\
$^{2}$ Physics Department, Princeton University, Princeton, New Jersey
08540\\
$^{3}$ Fundamental Research Laboratories, NEC Corporation, 34
Miyukigaoka, Tsukuba, Ibaraki 305-8501, Japan\\}
\date{\today}
\maketitle
\begin{abstract}
Experiments on quantum point contacts have
highlighted an anomalous conductance plateau around $0.7 (2e^2/h)$,
with features suggestive of the Kondo effect. Here we
present an Anderson model for transport through a point contact
which we analyze in the Kondo limit. Hybridization to the
band increases abruptly with energy but decreases with valence,
so that the background conductance and the Kondo temperature
$T_K$ are dominated by different valence transitions.
This accounts for the high residual conductance above $T_K$.
The model explains the gate-voltage, temperature, magnetic-field, and bias-voltage
dependence observed in the experiments.
A strongly spin-polarized current is predicted for Zeeman splitting
$g^* \mu_B B > k_B  T_K,k_BT$.
\end{abstract}
\begin{multicols}{2}

The conductance through quantum point contacts (QPCs) is observed
to be quantized in units of $2e^2/h$ \cite{vanWees,Wharam}. In
addition to these integer conductance steps, an extra conductance
plateau around $0.7 (2e^2/h)$ has attracted considerable
experimental effort \cite{Thomas,Kristensen,Reilly,hashimoto}
 and drawn attention to the effects of
electron-electron interaction on the transport properties of
low-dimensional quantum systems\cite{schmeltzer,Wang98,Spivak,Bruus}.
Interaction effects in QPCs may enable novel applications such as
solid-state spin filters\cite{Divincenzo}, detection of a single
charge and spin\cite{Engel}, and measurement and
read-out\cite{Aleiner,Gurvitz} of entangled spin states
in quantum information devices\cite{Loss}.

A recent experiment\cite{Cronenwett02}
has highlighted features in QPC transport
strongly suggestive of the Kondo effect: a zero-bias peak
in the differential conductance which splits in a magnetic
field, and a crossover to perfect transmission
below a characteristic ``Kondo" temperature $T_K$, consistent with
the peak width.  A puzzling observation was the large value of the
residual conductance, $G > 0.5 (2e^2/h)$, for $T \gg  T_K$.

Here we demonstrate the  applicability of an Anderson model to
transport through a QPC by comparing the results of perturbation
theory  in the Kondo limit  to experimental data. A novel feature
in the model distinguishes transport through a QPC from transport
through other Kondo impurities, {\it e.g.} quantum
dots\cite{Goldhaber}, and explains the large residual conductance:
the hybridization to the band is a strong function of energy and
valence. Predictions of the model include binding of an electron
at the QPC {\it before} the first conductance step, and a strongly
spin-polarized current at magnetic fields satisfying $g^* \mu_B B
> k_B T_K,k_B T$.

Use of an Anderson model for a QPC is motivated below by
 spin-density-functional-theory results indicating that a single
electron can bind at the center of the QPC. An intuitive picture
is to consider transport across a square barrier. For a wide and
tall barrier, in addition to the exponentially increasing
transparency, there are  narrow transmission resonances above the
barrier. These result from
 multiple reflections from the edges of the barrier, and
 are associated
with quasi-bound states,  which
 can play the role of localized orbitals in an Anderson
model. Our SDFT results indicate that even an initially smooth
QPC potential can produce a narrow quasi-bound
state, resulting in a spin bound at the center of the QPC.
We thus model the QPC and its leads by the Anderson
Hamiltonian\cite{Anderson}
\begin{eqnarray}
 H &=& \sum_{\sigma;k{\in L,R} }\!\!\! \e_{k\sigma}
{\bf c}_{k\sigma}^{\dagger} {\bf c}_{k\sigma}
  + \sum_\sigma\! \e_{\sigma}
{\bf d}_{\sigma}^{\dagger} {\bf d}_{\sigma}
  +  U {\bf n}_\uparrow {\bf n}_\downarrow \nonumber\\
  &+& \!\!\!\sum_{\sigma;k{\in L,R} }\!\! [V_{k\sigma}^{(1)}
      (1 - {\bf n}_{\bar{\sigma}})
     {\bf c}_{k\sigma}^{\dagger} {\bf d}_{\sigma}
                            + V_{k\sigma}^{(2)}
       {\bf n}_{\bar{\sigma}}
     {\bf c}_{k\sigma}^{\dagger} {\bf d}_{\sigma}
     + H.c.]
\label{HA}
\end{eqnarray}
where ${\bf c}_{k\sigma}^{\dagger} ({\bf c}_{k\sigma})$ creates
(destroys) an electron with momentum $k$ and spin $\sigma$ in one
of the two leads $L$ and $R$, ${\bf d}_{\sigma}^{\dagger} ({\bf
d}_{\sigma})$ creates (destroys) a spin-$\sigma$ electron on ``the
site", {\it i.e.} the quasi-bound state at the center of the QPC,
and ${\bf n}_{\sigma} = {\bf d}_{\sigma}^{\dagger} {\bf
d}_{\sigma}$. The hybridization matrix elements,
$V_{k\sigma}^{(1)}$ for transitions between 0 and 1 electrons on
the site and $V_{k\sigma}^{(2)}$ for transitions between 1 and 2
electrons, are taken to be step-like functions of energy,
mimicking the exponentially increasing transparency 
(the position of the step defines our zero of energy).
Physically, we expect $V_{k\sigma}^{(2)} < V_{k\sigma}^{(1)}$, as
the Coulomb potential of an electron already occupying the QPC
will reduce the tunneling rate of a second electron through the
bound state. In the absence of magnetic field the two spin
directions are degenerate,
$\e_\downarrow=\e_\uparrow=\e_0$.

For a noninteracting system, the  conductance $G$ will be a
(temperature broadened) resonance of Lorentzian form, with a width
proportional to $V^2$. If $V$ rises abruptly to a large value,
such that the width becomes larger than $\e_F-\e_0$, where $\e_F$
is the Fermi energy, $G$ saturates to a value of $2e^2/h$. For the
interacting system, we similarly expect the {\it high-temperature}
contribution from the $0\leftrightarrow1$ valence fluctuations to
$G$ to saturate  at $0.5 (2e^2/h)$ for $\e_F>0>\e_0$, because the
probability of an opposite spin electron occupying the site in
this regime is $\approx0.5$.
%
Since $V_{k\sigma}^{(2)}$ may be significantly smaller than
$V_{k\sigma}^{(1)}$, the contribution to the conductance from the
$1\leftrightarrow2$ valence fluctuations may be small, until $\e_F
\simeq \e_0+U$. However, the Kondo effect will enhance this
contribution with decreasing temperature, until at zero
temperature the conductance will be equal to $2e^2/h$, due to the
Friedel sum rule \cite{langreth} for the Anderson model.

To obtain a quantitative estimate of the conductance we
note that the relevant gate-voltage range corresponds to the
Kondo regime (singly occupied site), a fact further supported by
the observation of a zero-bias peak where the conductance first
becomes measurable \cite{Cronenwett02}, so the Kondo limit of the
Anderson Hamiltonian should be applicable. We therefore perform a
Schrieffer-Wolff transformation\cite{Schrieffer} to obtain the
Kondo Hamiltonian \cite{Kondo,RG}
\begin{eqnarray}
 H &=& \sum_{\sigma;k{\in L,R} }\!\!\!\!\!\e_{k\sigma}
{\bf c}_{k\sigma}^{\dagger} {\bf c}_{k\sigma}
 + \!\!\!\!\!\!\!\!\!\!\sum_{\sigma, \sigma';k,k' \in L,R}\!\!\!\!\!\!\!\!\!\!
    (J^{(1)}_{k k'\sigma \sigma} - J^{(2)}_{k k'\sigma \sigma} )
\,{\bf c}_{k\sigma}^{\dagger} {\bf c}_{k'\sigma} \nonumber \\
 &+&\  2\!\!\!\!\!\!\!\!\!\!\!\!\!\!\!
\sum_{\sigma, \sigma', \alpha, \alpha';k,k' \in L,R}\!\!\!\!\!\!\!\!\!\!\!\!\!
 (J^{(1)}_{k k'\sigma \bar{\sigma}} + J^{(2)}_{k k'\sigma \bar{\sigma}} )
 ( {\bf c}_{k\sigma}^{\dagger}
       \vec{\sigma}_{\sigma \sigma'} {\bf c}_{k'\sigma'}) \cdot
       \vec{S}, \\ \nonumber
&J&\!\!^{(i)}_{k k'\sigma \sigma'} = {{(-)^{i+1}}\over 4}\,
  \left( {{{V^{(i)}_{k\sigma} V^{*(i)}_{k'\sigma'}}}
   \over {\e_{k \sigma} - \e^{(i)}_\sigma} }
     + {{{V^{(i)}_{k\sigma} V^{*(i)}_{k'\sigma'}}
      \over {\e_{k' \sigma'} - \e^{(i)}_{\sigma'}}}}\right),
\label{HK}
\end{eqnarray}
where $\e^{(1)}_{\sigma} = \e_{\sigma}$ and $\e^{(2)}_{\sigma} =
\e_{\sigma} + U$. The Pauli spin matrices are indicated by
$\vec{\sigma}$, and the local spin due to the bound state  is
$\vec{S}\equiv{1 \over 2}{\bf d}_{\alpha}^{\dagger}
\vec{\sigma}_{\alpha \alpha'} {\bf d}_{\alpha'}$.

Following Appelbaum \cite{Appelbaum},  we treat the above Kondo
Hamiltonian perturbatively in the couplings $J^{(i)}_{k k'\sigma
\sigma'}$.
The differential conductance to lowest order, $J^2$, is
given by
\begin{eqnarray}
G_2 &=& {{4 \pi e^2}\over \hbar} \rho_L({\e_F}) \rho_R({\e_F})
\left\{(J^{(-)}_{LR})^2 + (J^{(+)}_{LR})^2  \right. \nonumber \\
&\times& \left. \left[\, 3 +2\langle M\rangle \left(\tanh{{\Delta+ eV}\over{2 k_B
T}} +  \tanh{{\Delta- eV}\over{2 k_B T}}\right) \right] \right\}
\label{Gtwo}
\end{eqnarray}
where, for simplicity, $J^{(i)}_{k k'\sigma \sigma'}$ are replaced by their
(magnetic-field independent) values
 at the Fermi energy
\begin{equation}
J^{(i)}_{l\,l'}
   \equiv J^{(i)}_{k_F \in l \ k_F \in l'\  \sigma
   \sigma}={{(-)^{i+1}V_i^2}
\over{ 2(\e_F-\e_0^{(i)} )}} f_{FD}(-\e_F/\delta) , \label{Jstep}
\end{equation}
where symmetric leads have been assumed, and the $V_i$ and
$\delta$ are constants. The $J^{(i)}$  increase in a step of the
Fermi-Dirac form $f_{FD}(x)=1/[1 + \exp(x)]$. We define the
combinations $J^{(\pm)}_{l\,l'} = J^{(1)}_{l\,l'}\, \pm\,
J^{(2)}_{l\,l'}$ for, respectively, the direct and exchange
couplings in Eq.~(2). In (\ref{Gtwo}),
 $\Delta = g^* \mu_B B$ is the Zeeman splitting,
$\langle M \rangle = -(1/2) \tanh(\Delta/2k_B T)$ is the
magnetization for the uncoupled site, and $\rho_{L/R}(\e) =
\sum_{k \in L/R} \delta(\e - \e_{k\sigma})$ is the single-spin
electron density of states in the leads. We assume $\rho =
\rho_L(\e) = \rho_R(\e)$.

As low temperatures the Kondo effect leads to a logarithmically
diverging contribution $G_3$ ({\it cf.} \cite{Appelbaum}) to the
differential conductance at order $J^3$ \cite{Kondo}, due to
integrals running from the Fermi energy to either band edge.
Because of the steplike increase of the $J^{(i)}$, the band
integral for $J^{(1)}$ runs down from $\e_F$ to the hybridization
step at zero, but runs up from $\e_F$ to $\e_0 + U$ for $J^{(2)}$.
Since in the region of interest $\e_0 + U - \e_F \gg \e_F$, {\it
the logarithmic contribution from} $J^{(2)}$ {\it dominates}
$G_3$.

\vskip 1.5 truecm
\begin{center}
\leavevmode \epsfxsize=4in \epsfbox[62 -98 626 366]{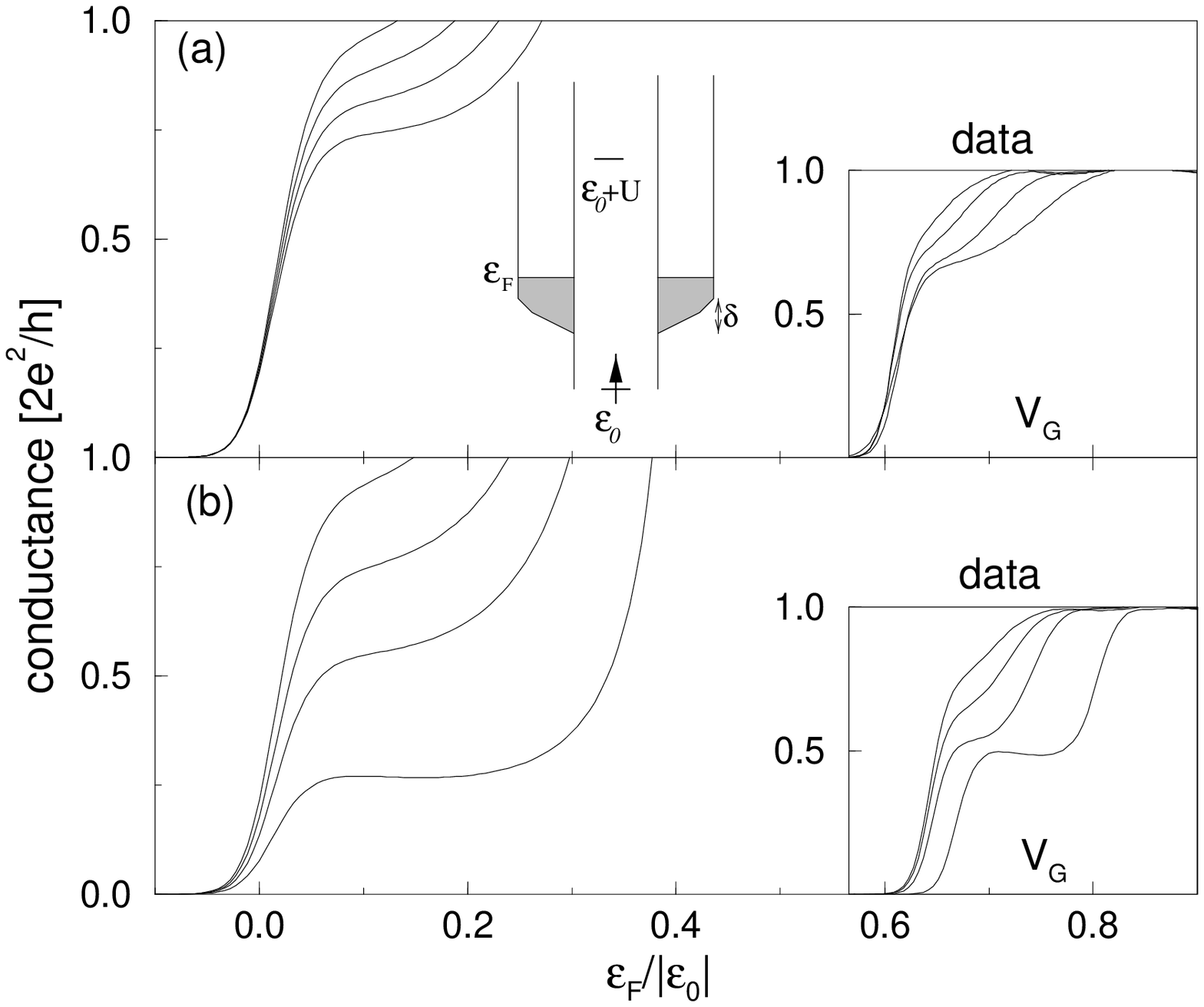}
\end{center}
\begin{small}
\vskip -3 truecm Fig.1: Results of the Kondo model. (a)
Conductance at temperatures $T=0.05,0.1,0.2,0.6$, (solid curves,
from high to low) as a function of Fermi energy $\e_F$ (all
energies in units of $|\e_0|$). The parameters are $U=1.45,\rho
V_1^2=0.12,\rho V_2^2=0.015$ and $\delta=0.02$. Right inset:
experimental conductance of QPC at
4 different temperatures \cite{Cronenwett02}. Center inset:
Schematic of the band structure for our Anderson model. (b)
Conductance in a magnetic field, for Zeeman splitting
$\Delta=0,0.07,0.12,0.4$
 at $T=0.06$ (solid curves from top to bottom).
  Inset: experimental conductance
of QPC at different magnetic fields
 \cite{Cronenwett02}.
\end{small}
\vskip 0.2 truecm

Fig.~1 depicts the linear-response conductance ($G_2+G_3$). Since
$G_2$ depends only on the values of $J^{(i)}_{LR}$ at $\e_F$, it
is dominated by $J^{(1)}$, while the Kondo enhancement is
dominated by $J^{(2)}$. As argued above, the contribution due to
$J^{(1)}$ is set around $0.5 (2e^2/h)$ by construction,  while the
contribution due to $J^{(2)}$, resulting from the
$1\leftrightarrow2$ valence fluctuations is small at high
temperature, but grows with decreasing temperature in a form
following the Kondo scaling function, $F(T/T_K)$, where $T_K
\simeq U \exp(-1/4\rho J^{(2)})=U \exp[(\e_F-\e_0-U)/2\rho V_2^2]$,
in agreement with the experimental observation of a Kondo
temperature increasing exponentially with gate voltage $\sim
\e_F$. Note that in perturbation theory the conductance is not
bound by its physical limit: $2e^2/h$.

The dependence of conductance on magnetic field is shown in
Fig.~1(b). The Kondo logarithms in $G_3$ are suppressed and the
term in $G_2$ that depends on $\langle M\rangle$ gives a negative
contribution $\propto \tanh^2(\Delta/2k_B T))$, leading to the
evolution of the $0.7$ plateau towards and below $0.5$. In
agreement with experiment\cite{Cronenwett02}, the conductance is
no longer monotonically increasing with Fermi energy $\e_F$: the
energy denominator causes the $J^{(1)}$ contribution to $G_2$ to
decrease, and this is no longer compensated by an increase of
$G_3$. Due to shortcomings of perturbation theory the conductance
at large magnetic field reduces to a value smaller than $0.5
(2e^2/h).$

\vskip 1.5 truecm
\begin{center}
\leavevmode \epsfxsize=4in
\epsfbox[62 -98 626 366]{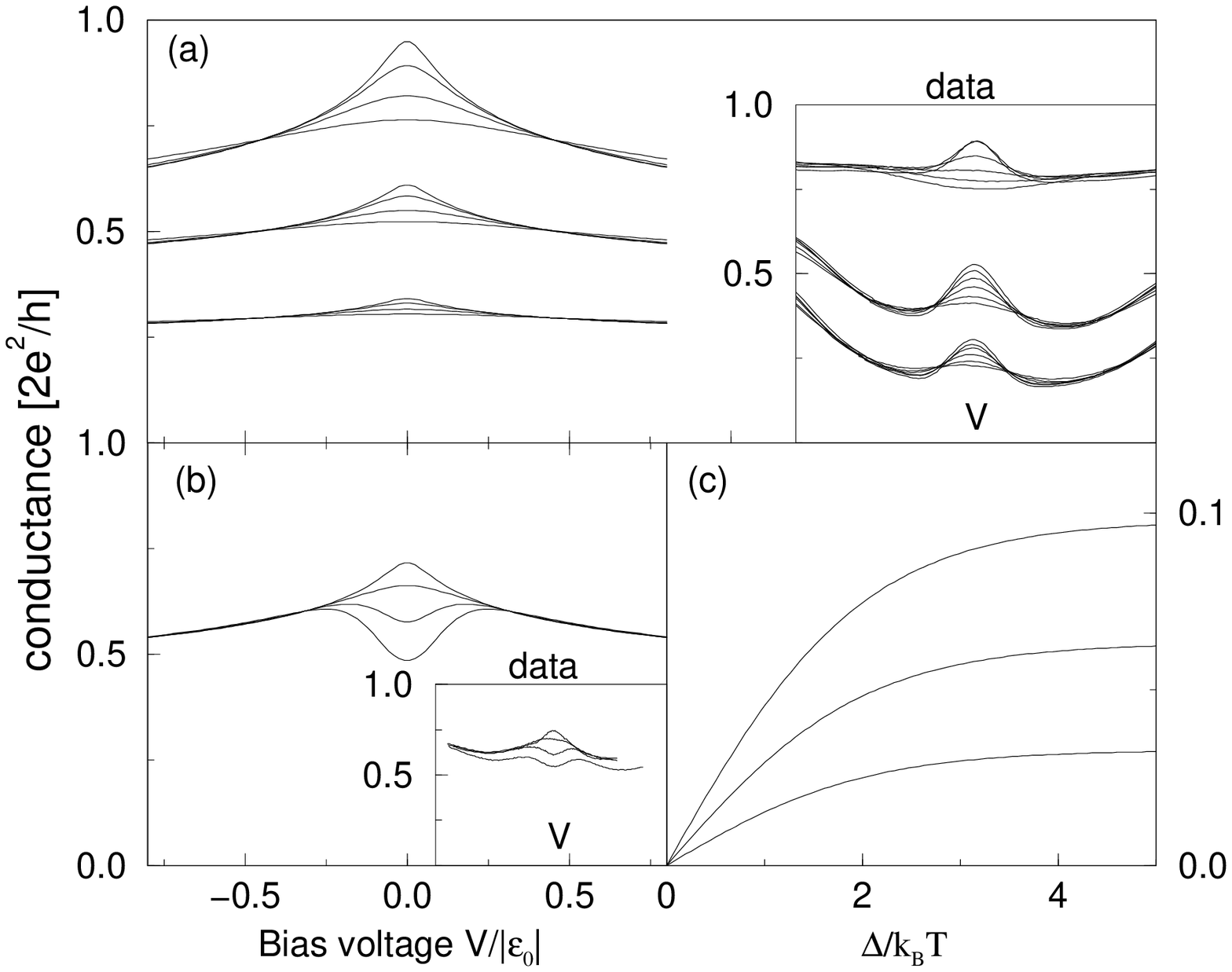}
\end{center}
\begin{small}
\vskip -3 truecm Fig.2: Differential conductance $dI/dV$ for the
Kondo model. (a) $dI/dV$ versus bias at Fermi energies $\e_F =
0.1, 0.03, 0.01$ from the top group to the bottom. For each
chemical potential curves are shown for temperatures
$T=0.06,0.1,0.2,0.4$ from top to bottom. All other parameters are
the same as in Fig.~1.
Inset: experimental differential conductance \cite{Cronenwett02}.
(b) $dI/dV$ in magnetic fields with Zeeman splitting
$\Delta=0,0.04,0.07,0.1$ at $T= 0.06$ and $\mu=0.04$. Inset:
experimental differential conductance at different magnetic fields
\cite{Cronenwett02}. (c) Spin conductance $d(I_\downarrow -
I_\uparrow)/dV$ as a function of magnetic fields, for several
values of $\e_F$.
\end{small}
\vskip 0.2 truecm

Fig.~2(a) shows the differential conductance as a function
of bias voltage,
for several values of $\e_F$ and temperatures.
Even at the lowest
conductances (small $\e_F$) there is a clear Kondo
peak, as is seen in experiment (inset). Due to the suppression
 of the
Kondo effect by voltage, the large voltage traces are independent
of temperature, again in agreement with experiment. Magnetic field
splits the Kondo peak as shown in Fig.~2(b).

An important prediction of the Kondo model is that the current
through a QPC will be spin polarized if the Zeeman splitting is
larger than both $k_B T$ and $k_B T_K$ (Fig.~2(c)). The net spin
conductance $G_\sigma$, is given, to second order in $J$, by
\begin{equation}
G_\sigma = {{16 \pi e^2}\over \hbar} \rho^2 \langle M \rangle
\left[ (J_{LR}^{(1)})^2-(J_{LR}^{(2)})^2\right] .
\label{Gs}
\end{equation}
Therefore, at low temperatures and in the vicinity of the $0.7
(2e^2/h)$ plateau where $T_K$ is small, a QPC can be an effective
spin filter at weak magnetic fields ($\Delta > k_B T_K,k_B T$).

Lastly, we present evidence from spin-density-functional
theory (SDFT) \cite{HK} for the formation of a local moment (bound spin) at
the center of a GaAs QPC, which supports our use of the Anderson
model. SDFT is applied within the local-density
approximation\cite{Kohn,Callaway}.
The external potential consists of a clean quantum wire
with a parabolic confining potential
of $V_{\rm wire}^{0}(y)=(1/2)m^{*}\omega_y^2 y^2$ and a QPC potential
\begin{equation}
V_{\rm QPC}(x,y)=V(x)/2 +
m^{*}\left[V(x)/\hbar\right]^2 y^2/2,
\end{equation}
where $V(x) = V_0/\!\cosh^2(\alpha x)$,
with $\alpha=\omega_x\sqrt{m^{*}/2V_0}$. A contour plot of the QPC
potential $V_{\rm QPC}(x,y)$ is shown in the left inset
of Fig.~3(b).

We solve the
Kohn-Sham equation\cite{Kohn} using the material constants for GaAs,
 $m^*=0.067m_0$ and
$\kappa=13.1$.  
The external confinement in the $y$-direction in the wire is fixed by
$\hbar\omega_y=2.0{\rm meV}$.
The parameters for the QPC potential are taken to be
$V_0=3.0{\rm meV}$ and $\hbar\omega_x=1.5{\rm meV}$.

\
Fig.~3(a) shows the spin-dependent, self-consistent QPC barriers
at $T$=0.1K obtained from SDFT \cite{later}. Specifically, we plot
the energy of the bottom of the lowest 1D subband
$\epsilon_\sigma(x)$, relative to the value $\epsilon_0$ far into
the wire, for both spin-up and spin-down. The local density of
states $\nu(\epsilon)$ at the center of the QPC is shown for both
spin-up and spin-down in the right inset. Fig.~3(b) shows the
average 1D electron density through the QPC and the net density of
spin-up electrons. The integrated spin-up density is 0.96
electrons. The data from SDFT gives strong evidence for a
quasi-bound state centered at the QPC: there is a resonance in the
local density of states $\nu(\epsilon)$ for spin-up, with a net of
one spin bound in the vicinity of the QPC.  The transmission
coefficient $T(\epsilon)$ for electrons in the lowest subband is
shown in the left inset to Fig.~3(a). Transmission for spin-up is
approximately 1 over a broad range of energies above the spin-up
resonance. This implies an onset of strong
hybridization 
at energies above the quasi-bound
state.

We have presented a microscopic Anderson model, supported by
spin-density-functional theory, for transport through a quantum
point contact. The anomalous $0.7 (2e^2/h)$ plateau is attributed
to a high background conductance plus a Kondo enhancement. The
temperature scales for these two contributions are decoupled:
$0 \leftrightarrow 1$ valence transitions account for the
background conductance, while $1 \leftrightarrow 2$ valence
transitions give the dominant Kondo effect. Based on this model
one can make specific experimental predictions. A strongly
spin-polarized
 current is predicted when the Zeeman
splitting exceeds both $k_B T$ and $k_B T_K$. The predicted
formation of a bound state (local moment) can be directly tested
by measuring transport through two parallel point contacts,
coupled capacitively, with one of them tuned to  $G\simeq e^2/h$,
{\it i.e.} in the region of maximal sensitivity to its
environment. When the gate voltage controlling the other point
contact is scanned through the electron binding event
 (predicted to occur for $G\ll e^2/h$), an abrupt
decrease should be seen in the conductance of the half-transparent
point contact. (A very similar arrangement was used recently to
probe the bound states of a quantum dot \cite{sprinzak}.) The
presence of bound spins in QPCs near pinch-off has potentially
profound effects on transport through quantum dots with QPCs as
leads. In particular the leads may act as magnetic impurities, and
cause the apparent saturation of the dephasing time in transport
through open semiconductor quantum dots at low temperatures
\cite{dephasing}, and may complicate attempts to measure the spin
of dot electrons.\cite{folk}.

\vskip -1.5 truecm
\begin{center}
\leavevmode \epsfxsize=4.5in \epsfbox[112 222 676 686]{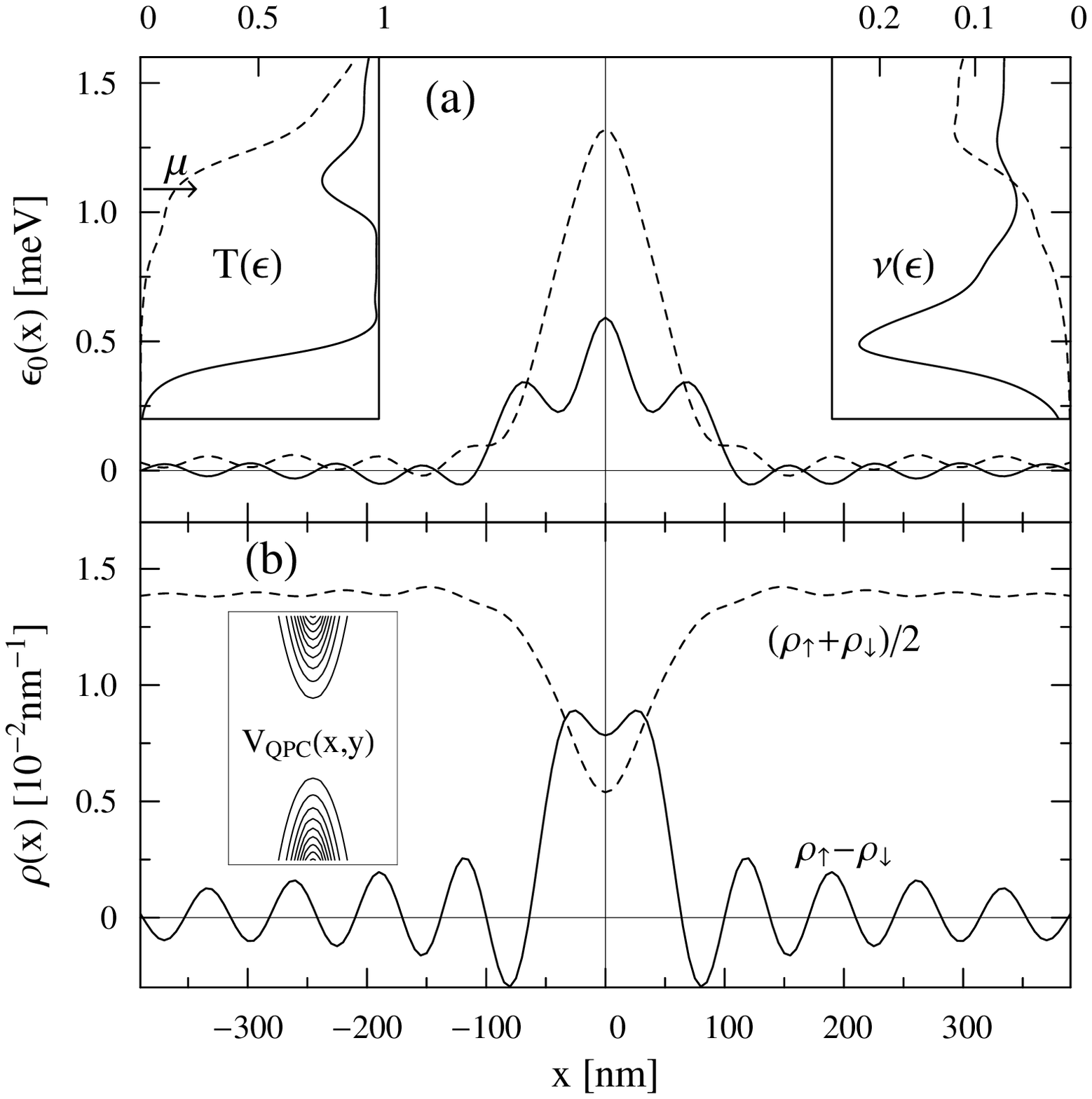}
\end{center}
\begin{small}
\vskip -1.0 truecm
Fig.3: Results of spin-density-functional theory.
(a) Self-consistent ``barrier", {\it i.e.} energy of the
bottom of the lowest 1D subband at temperature $T$ = 0.1K
as a function of position $x$ in the direction of current
flow through the QPC. The electrochemical potential $\mu$
is indicated by an arrow on the left.
In this panel, solid curves are for spin-up electrons
and dashed curves are for spin-down electrons.
Left inset: transmission coefficient.
Right inset: local density of states at center of QPC.
(b) 1D electron density in QPC. The solid curve gives
the net spin-up density and the dashed curve gives
the spin-averaged density. Inset: contour plot of the
QPC potential $V_{\rm QPC}(x,y)$.
\end{small}
\vskip 0.2 truecm

The calculations presented in this paper were perturbative and
thus the comparison with experiment could only
besemi-quantitative. The main failure of perturbation theory is
its inability to obtain the low-temperature unitarity limit
$2e^2/h$. We hope that our work will motivate more accurate
treatments of the Anderson and Kondo models introduced here.

We acknowledge fruitful discussions with B.~Altshuler,
D.~Goldhaber-Gordon, B.~Halperin, C.~Marcus, and M.~Pustilnik.
Y.M. acknowledges partial support by NSF grant DMR 00-93079.

$^\dagger$ Permanent Address:
Department of Physics, Ben-Gurion University, Beer Sheva
84105, Israel.

\end{multicols}

\begin{references}

\bibitem{vanWees} B.~J.~van~Wees {\it et al.},
{\sl Phys.~Rev.~Lett.} {\bf 60}, 848 (1988).
\bibitem{Wharam} D.~A.~Wharam {\it et al.},
{\sl J.~Phys.~C} {\bf 21}, L209 (1988).


\bibitem{Thomas} K.~J.~Thomas {\it et al.},
{\sl Phys.~Rev.~Lett.} {\bf 77}, 135 (1996);
K.~J.~Thomas {\it et al.},
{\sl Phys.~Rev.}~{\bf B58}, 4846 (1998).

\bibitem{Kristensen} A.~Kristensen {\it et al.},
{\sl Phys.~Rev.} {\bf B62}, 10950 (2000).

\bibitem{Reilly} D.~J.~Reilly {\it et al.},
{\sl Phys. Rev.} {\bf B63}, 121311 (2001).

\bibitem{hashimoto}
S. Nuttinck {\it et al.}, {\sl
Jap. J. of App. Phys} {\bf 39}, L655 (2000);
K. Hashimoto {\it et al.}, {\sl
ibid} {\bf 40}, 3000 (2001).

\bibitem{schmeltzer}
D. Schmeltzer {\it et al.}, {\sl
Phil. Mag.}  {\bf B 77}, 1189 (1998).

\bibitem{Wang98} C.-K.~Wang and K.-F.~Berggren,
Phys.~Rev. {\bf B57}, 4552 (1998).

\bibitem{Spivak} B.~Spivak and F.~ Zhou,
{\sl Phys.~Rev.} {\bf B61}, 16730 (2000).

\bibitem{Bruus} H.~Bruus, V.~V.~Cheianov, and K.~Flensberg,
{\sl Physica}~{\bf E 10}, 97 (2001).

\bibitem{Divincenzo} D.~P.~Divincenzo, {\sl J.~Appl.~Phys.} {\bf 85}, 4785
(1999).

\bibitem{Engel} H.-A.~Engel and D.~Loss, {\sl Phys.~Rev.~Lett.} {\bf 86}, 4648
(2001).
\bibitem{Aleiner} I.~L.~Aleiner, N.~S.~Wingreen, and Y.~Meir,
{\sl Phys.~Rev.~Lett.} {\bf 79}, 3740 (1997).
\bibitem{Gurvitz} S.~A.~Gurvitz, {\sl Phys.~Rev.~Lett.} {\bf 85}, 812 (2000).
\bibitem{Loss} D.~Loss and D.~P.~Divincenzo, {\sl Phys.~Rev.} {\bf A57},
120 (1998).
\bibitem{Cronenwett02} S.~M.~Cronenwett {\it et al.}, {\sl Phys. Rev. Lett.}
{\bf 88}, 226805 (2002).
\bibitem{Goldhaber} D.~Goldhaber-Gordon {\it et al.}, {\sl Nature} (London)
{\bf 391}, 156 (1998).
\bibitem{Anderson} P.~W.~Anderson, {\sl Phys.~Rev.}~{\bf 124}, 41 (1961).
\bibitem{langreth}D. C. Langreth, {\sl Phys. Rev.} {\bf 150}, 516 (1966);
See also T.~K. Ng and P.~A. Lee, {\sl Phys. Rev. Lett.} {\bf 61}, 1768 (1988).
\bibitem{Schrieffer} J.~R.~Schrieffer and P.~A.~Wolff,
{\sl Phys.~Rev.} {\bf 149}, 491 (1966).
\bibitem{Kondo} J. Kondo, {\sl Prog. Th. Phys.} (Kyoto) {\bf 32}, 37 (1964).
\bibitem{RG} The parameters appearing in the Kondo Hamiltonian are not the
bare parameters of the Anderson model (\ref{HA}), but renormalized
parameters after the bandwidth has been reduced to $U$ [F.~D.~M.
Haldane, {\sl Phys. Rev. Lett.} {\bf 40}, 416 (1978)].
\bibitem{Appelbaum} J.~A.~Appelbaum, {\sl Phys.~Rev.} {\bf 154}, 633 (1967).
Appelbaum approximates the diverging integrals  by
$\log(|A|+k_bT)$. We use $\log[A^2+(k_bT)^2]$ instead.
\bibitem{HK}
P. Hohenberg and W. Kohn, {\sl  Phys. Rev.} {\bf 136}, B864 (1964).
\bibitem{Kohn} W.~Kohn and L.~J.~Sham, {\sl Phys.~Rev.} {\bf 140}, A1133
(1965). We use the local-density approximation for the
exchange-correlation energy
$E_{\rm xc}=\int\!\rho({\bf r})\,\e_{\rm xc}
[\rho({\bf r})]\, d{\bf r}$, where $\e_{\rm xc}[\rho({\bf r})]$
is the parameterized form by Tanatar and Ceperley for the
two-dimensional electron gas
[B.~Tanatar and D.~M.~Ceperley, {\sl Phys.~Rev.} {\bf B39}, 5005 (1989)].
\bibitem{Callaway} J.~Callaway and N.~H.~March, {\sl Solid State Phys.}
{\bf 38}, 135 (1984).


\bibitem{later}
The solution with broken spin-symmetry
coexists with an unpolarized solution
(K. Hirose, N. S. Wingreen, and Y. Meir, in preparation).
See also
A.~M. Bychkov, I.~I. Yakimenko, and K-F Berggren, {\sl Nanotechnology}
 {\bf 11}, 318 (2000).
\bibitem{sprinzak}
D. Sprinzak {\it et al.}, {\sl Phys. Rev. Lett.} {\bf 88}, 176805 (2002).
\bibitem{dephasing}
D.P. Pivin {\it et al.}, {\sl Phys. Rev. Lett.} {\bf 82}, 4687 (1999);
A.G. Huibers {\it et al.}, {\sl Phys. Rev. Lett.} {\bf 83}, 5090 (1999).
\bibitem{folk}
J. A. Folk {\it et al.}, {\sl Phys. Scr.} {\bf T 90}, 26 (2001).
\end{references}
\end{document}